\documentclass[final,5p,times,twocolumn]{elsarticle}

\usepackage{amssymb,amsmath,bm,gensymb,color}
\usepackage{graphicx,float,subfig}
\usepackage{lineno}

\usepackage{hyperref}

\hypersetup{
    colorlinks=true,
    linkcolor=black,
    citecolor=black,
    filecolor=black,
    urlcolor=black,
}

\biboptions{sort&compress}

\journal{International Journal of Solids and Structures}

\begin{document}

\begin{frontmatter}

\title{Temperature-induced shape morphing of bi-metallic structures}
\author[add1]{S. Taniker}
\author[add1]{P. Celli}
\author[add2]{D. Pasini}
\author[add3]{D.C. Hofmann\corref{cor1}}
\ead{dch@jpl.nasa.gov}
\author[add1]{C. Daraio\corref{cor1}}
\ead{daraio@caltech.edu}
\cortext[cor1]{Corresponding authors}

\address[add1]{Division of Engineering and Applied Science, California Institute of Technology, Pasadena, CA 91125, USA}
\address[add2]{Department of Mechanical Engineering, McGill University, Montreal, QC H3A 0C3, Canada}
\address[add3]{Jet Propulsion Laboratory/California Institute of Technology, Pasadena, CA 91109, USA}

\begin{abstract}
In this work, we study the thermo-mechanical behavior of metallic structures designed to significantly change shape in response to thermal stimuli. This behavior is achieved by arranging two metals with different coefficient of thermal expansion (CTE), Aluminum and Titanium, as to create displacement-amplifying units that can expand uniaxially. In particular, our design comprises a low-CTE bar surrounded by a high-CTE frame that features flexure hinges and thicker links. When the temperature increases, the longitudinal expansion of the high-CTE portion is geometrically constrained by the low-CTE bar, resulting in a large tangential displacement. Our design is guided by theoretical models and numerical simulations. We validate our approach by fabricating and characterizing individual units, one dimensional arrays and three-dimensional structures. Our work shows that structurally robust metallic structures can be designed for large shape changes. The results also demonstrate how harsh environmental conditions (e.g., the extreme temperature swings that are characteristic of extraterrestrial environments) can be leveraged to produce function in a fully passive way.

\vspace{15px}
\noindent{\textbf{This article may be downloaded for personal use only. Any other use requires prior permission of the authors and Elsevier Publishing. This article appeared in}: \emph{International Journal of Solids and Structures} {\bf 190}, 22--32 (2020) \textbf{and may be found at}: \url{https://doi.org/10.1016/j.ijsolstr.2019.10.024}}
\vspace{10px}
\end{abstract}


\end{frontmatter}


\section{Introduction}

In most engineering applications requiring extreme environments, structures have to be designed to resist deformations during large temperature variations. For example, components of internal combustion engines, biomedical devices and spacecraft can be exposed to extreme, yet predictable, temperature changes. The increasing need for new, light-weight, multifunctional materials has motivated studies on materials and structures with tailored coefficient of thermal expansion (CTE)~\cite{yoshida2004} and, especially, low or negative CTE. One way to obtain such properteis is by engineering materials with peculiar microstructures~\cite{slack1975,roy1989,mary1996,greve2010,das2010,grobler2013}. Another strategy consists of designing structures featuring arrangements of two or more materials with different CTEs~\cite{sigmund1996,lakes2007,jefferson_tailorable_2009,wei2016jmps,wei2018,li_hoberman-sphere-inspired_2018}. For example,creating multi-layer solids with zero CTE~\cite{wetherhold1995} and architected  unit cells arranged in periodic media ~\cite{schaedler2016} with extreme thermal expansion~\cite{sigmund1996,sigmund1997}. Two-dimensional zero-CTE architected structures were then realized at the macroscale~\cite{steeves2007,berger2011,gdoutos2013,parsons2019}, and at the microscale~\cite{yamamoto2014}, to create thin thermally-stable films that are potentially applicable as space telescope mirrors. Recently, others have extended this paradigm to hierarchical  arrangements  with theoretically unbounded thermal expansion~\cite{xu2017multilevel} and three-dimensional structures with tailorable CTE, including  zero or negative~\cite{xu2016,xu2018,wang2016,qu2017,heo2017} values. 

 In recent years, there have been also a few demonstrations of structures that can leverage temperature changes to attain mechanical motion. A typical example is that of bi-metallic beams and plates that can significantly change shape when heated. For example, bi-metal springs have been used to passively actuate louvers for the thermal regulation of spacecrafts~\cite{gilmore2002}. Recently, the use of bi-metallic cantilevers has been used to achieve similar motion, both in the context of space structures~\cite{athanasopoulos2017} and building envelopes or fa\c{c}ades~\cite{sung2016}. Moving beyond simple bilayers, which restrict the available shapes and structural responses that can be achieved, others have explored more complex architectures with the same goal of obtaining large displacements. Examples include origami~\cite{boatti2017}, kirigami~\cite{tang2017,cui_pop-up_2017}, metamaterials featuring bi-layer faces and connections, knitted arrays of thermally responsive fibers~\cite{han_blooming_2017}, as well as bi-material lattices capable of extreme shape changes~\cite{pasini2019}. The combination of architecture and material properties can allow  significant shape morphing in response to temperature variations. For example, structures composed of liquid crystal elastomers~\cite{kotikian2018} or shape memory materials~\cite{hartl2007,peraza2013,yuan2017} benefit from the temperature responsiveness and large deformability of their constitutive elements.  However, adapting these architectures to typical structural materials (e.g., metals) is quite challenging, since the CTE of metals is smaller than those found, for example, in LCE polymers. In addition, the elastic-plastic behavior common in metallic elements poses additional challenges when designing structures undergoing large deformations.

Designing metallic structures capable of large thermal expansion involves engineering mesoscale architectures that can leverage the small thermal strains typical of metals ($\sim \mathrm{10^{-6}/\degree C}$) to produce large global displacements. In other contexts, e.g., to amplify piezoelectric strains~\cite{idogaki_bending_1996,ervin_recurve_1998,jaenker_electrostrictive_2001,kim_development_2004,juuti_mechanically_2005,muraoka_displacement_2010,york2017}, to harvest energy~\cite{feenstra_energy_2008,clingman_broadband_2008} or attenuate vibrations~\cite{acar2013,taniker2015,taniker2017}, a similar goal has been achieved using displacement amplification mechanisms. Typically, these mechanisms comprise a combination of rigid links and compliant hinges designed to leverage geometric constraining. They can lead to very large displacement amplification factors (e.g., $25\times$ ~\cite{juuti_mechanically_2005}) and can be also engineered to obtain bending motion~\cite{idogaki_bending_1996}. In order to increase the overall stroke, researchers have used arrays of displacement-amplifying units~\cite{idogaki_bending_1996,ervin_recurve_1998}. It is to be noted that few have also explored these mechanisms in the context of thermal displacements but, typically, at the microscale~\cite{howell2006,wu2018} or relying on bi-metallic flexures that are challenging to fabricate~\cite{heo2017,heo2019}.

In this study, we couple two metals with relatively high and low thermal expansion to create two- and three-dimensional structures capable of shape morphing, i.e., of achieving large overall deformations. Our two-dimensional displacement amplifying unit stems from the idea that the longitudinal motion of the high-CTE outer frame is constrained by the low-CTE inner bar. This, in turn, causes the outer frame to expand in the transverse direction when the structure is heated ~\cite{pasini2019}.We consider a compliant mechanism-inspired design that is suitable to work with metals that are typically required for spacecraft and planetary landers.
Provided that the out-of-plane thickness of our structure is  sufficiently large, the tangential expansion due to thermal expansion will only occur along the desired, in-plane direction. If we resort to an Aluminum-Titanium material couple (Al-6061-T6 and Grade 5 Ti-6Al-4V), our structure has the ability to almost double its width in response to a temperature change of $100\degree \mathrm{C}$. This bi-material selection and temperature range has direct application to the lunar surface, where structures have historically been fabricated from aluminum and titanium and where a predictable temperature swing of a few hundred degrees occurs between lunar day and night. We begin with a detailed numerical and theoretical  analysis of a single unit with a particular focus on the analysis of the compliant hinges of the unit. Our theoretical  model accounts for the elasticity of the structure and allows to probe in detail the influence of all design parameters. We then proceed to experimentally test micro-waterjet-manufactured arrays of ten units, and three-dimensional structures obtained by assembling multiple arrays. Our work provides insight into the realization of macroscopic, shape-morphing metallic structures that can leverage extreme environmental conditions to passively perform desired functions. This is particularly appealing to create low-part-count space structures, e.g., passive switches and deployable solar arrays for lunar applications, and passive louvers for thermal regulation. While our structures are assemblies of waterjet-cut parts, we envision the possibility of manufacturing them as a single part via direct energy deposition additive manufacturing technologies. 

This article is organized as follows. In Section~\ref{s:Num}, we describe our structure, its design parameters and its numerical response. In Section~\ref{s:Analy}, we introduce and compare a purely-kinematic model and a mechanics-based model, and use the latter to study the influence of the design parameters on the structure expansion. In Section~\ref{s:Exp}, we experimentally validate our results for a Ti-Al material couple. In Section~\ref{s:lun}, we modify our design to create structures that can morph during the lunar day-night cycle without undergoing plastic deformation. The conclusions of our work are drawn in Section~\ref{s:Concl}.

\section{Unit cell design and finite element analysis}
\label{s:Num}
\begin{figure}[!htb]
\centering
\includegraphics[scale=1]{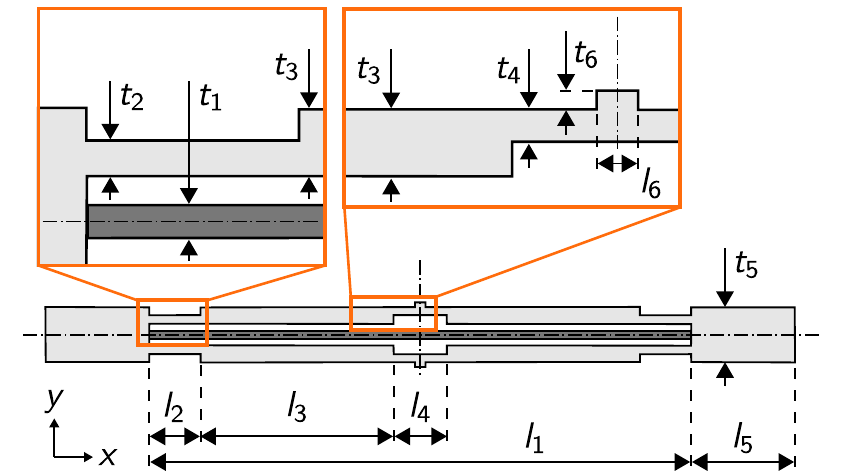}
\caption{Sketch of the displacement-amplifying unit analyzed in our finite element simulations, with its characteristic dimensions. The drawing is not on scale; refer to the text for the correct dimensions. The structure is symmetric about the dash-dot axes. The light gray outer frame is the high-CTE part; the inner dark gray bar is the low-CTE part.}
\label{f:NumericalUnit}
\end{figure}

A sketch of our unit, which represents the building block we use to create displacement-amplifying structures, is shown in Fig.~\ref{f:NumericalUnit}.
It comprises a high-CTE frame (light gray shaded area) and a low-CTE bar (dark gray shaded area). The two parts are assumed to be perfectly bonded at their interface. This design is inspired by a geometry for displacement amplification introduced in the context of vibration control~\cite{acar2013}, and it essentially differs from it for the addition of the low-CTE bar. This inner bar (hereby called beam 1) has length $l_1$ and thickness $t_1$. The high-CTE part comprises beams of length $l_3$ and thickness $t_3$ (beams 3) connected by flexure hinges of length $l_4$ and thickness $t_4$ (beams 4). Beams of type 3 are connected to the low-CTE bar via high-CTE blocks and flexure hinges of length $l_2$ and thickness $t_2$ (beams 2). In the following, unless otherwise specified, we set the following \emph{default} geometrical parameters: $l_2=l_4=6\,\mathrm{mm}$, $t_2=t_4=1\,\mathrm{mm}$, $l_3=79\,\mathrm{mm}$, $t_1=t_3=2\,\mathrm{mm}$ and $l_1=2l_2+2l_3+l_4$. Note that the size of the high-CTE blocks at the left and right edges of the unit does not play a major role in its thermomechanical response; we set $l_5=12\,\mathrm{mm}$ and $t_5=t_1+2t_3+2t_g$, where $t_g=0.25\,\mathrm{mm}$ is the thickness of the gap between beams 1 and 2. Finally, we add small connectors on beams 4, to facilitate the connection of one unit to others. These features have length $l_6=1.5\,\mathrm{mm}$ and thickness $t_6=0.375\,\mathrm{mm}$, and de facto reduce the length of the flexure hinges of type 4.

\begin{figure}[!htb]
\centering
\includegraphics[scale=1]{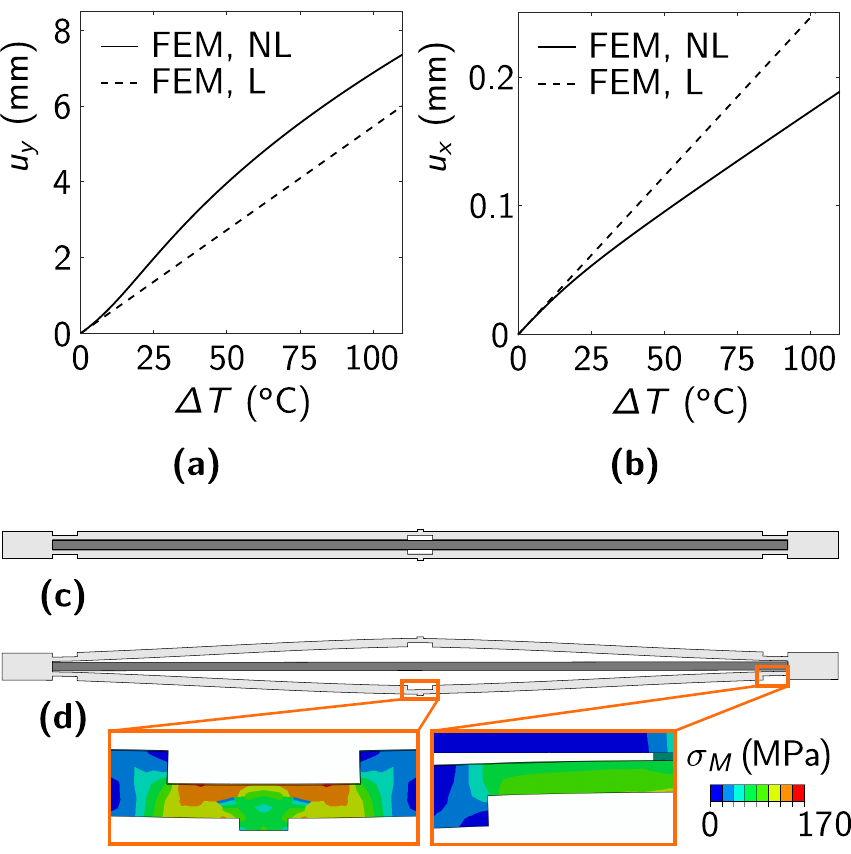}
\caption{(a) Total vertical displacement of a unit featuring the dimensions written in the main text, as a function of the temperature increment. The initial temperature for our simulation is $T_i=20 \degree \mathrm{C}$. {We compare results from simulations with and without geometric nonlinearities.} (b) Horizontal displacement of the unit as a function of the temperature increment. (c) Undeformed unit ($\Delta T= 0 \degree \mathrm{C}$). (d) Unit at $\Delta T=80 \degree \mathrm{C}$, with details illustrating the von Mises stress at two representative flexures.}
\label{f:NumericalResults}
\end{figure}

Our first step is to perform numerical simulations on this unit. To do so, we resort to the commercial finite element (FE) platform Abaqus/Standard. We consider a 3D model of our structure with depth $b=3\,\mathrm{mm}$. We select 20-node bi-quadratic elements. To produce large deformations in response to changes in temperature, we select materials with significantly-different CTE ($\alpha$). While combinations of polymers and polymers/metals offer the largest $\Delta \alpha$, we choose materials that are relevant for space structures. The inner, low-CTE bar is made of Titanium (Ti-6-4, $E_L=115\,\mathrm{GPa}$, $\nu_L=0.34$, $\alpha_L=8.6\,\,\mathrm{10^{-6}/\degree C}$) and the outer high-CTE frame is made of Aluminum (Al-6061-T4, $E_H=70\,\mathrm{GPa}$, $\nu_H=0.33$, $\alpha_H=23.1\,\,\mathrm{10^{-6}/\degree C}$)~\cite{ashby2016}. The behavior of these materials is assumed to be elastic-plastic with no strain hardening; the yield strength of the selected aluminum alloy is $\sigma_{Y}=260\,\mathrm{MPa}$, while that of titanium is $\sigma_{Y}=950\,\mathrm{MPa}$. We also assume the coefficients of thermal expansion to be constant over large temperature changes. The simulation results for a single unit are summarized in Fig.~\ref{f:NumericalResults}.
The initial temperature is $T_i=20 \degree \mathrm{C}$, while the final one is $T_f=T_i+\Delta T$. As we increase the temperature, the structure expands in the $y$ direction, as shown in the total vertical displacement plot of Fig.~\ref{f:NumericalResults}(a). In particular, we can appreciate that the compliant beams (\#2 and \#4) undergo quite large deformation due to geometric non linearity, which cannot be captured by linear FE analysis. At $\Delta T=80\,\mathrm{\degree C}$, we reach a total displacement of $5.8\,\mathrm{mm}$, corresponding to a vertical strain $\epsilon_{y}=0.8$. The expansion is almost uniaxial, and the horizontal displacement of the overall unit ($u_{x}$) is negligible with respect to the vertical one, as shown in Fig.~\ref{f:NumericalResults}(b). To visualize the deformation, we report the undeformed structure at $\Delta T= 0\,\degree \mathrm{C}$ and the expanded one at $\Delta T= 80\,\degree \mathrm{C}$ in Fig.~\ref{f:NumericalResults}(c) and (d), respectively. The details in Fig.~\ref{f:NumericalResults}(d) highlight the von Mises stress maps at representative flexure hinges; the maximum value ($\sigma_M\approx160\,\mathrm{MPa}$) is achieved at flexure 4, and it is of the same order of magnitude but lower than the yield strength of our aluminum of choice. This highlights that particular care needs to be placed in designing the compliant hinges to avoid plastic deformation for a prescribed temperature variation.

Later in this article, we analyze the behavior of stacks of units connected by means of the small connectors on beams of type 4 shown in Fig.~\ref{f:NumericalUnit}. Note that, in these cases, the expansion of an array of units scales linearly with the number of units in the array. This implies that we neglect gravity effects in our calculations.

\section{Theoretical analysis and influence of design parameters}
\label{s:Analy}
To better understand the influence of the design parameters on the structure response, without the need for numerical simulations, we develop simplified theoretical models. First, we treat our structure as a pin-jointed mechanism, by considering a purely-kinematic model. Afterwards, we introduce bending and stretching, to obtain a better approximation of the actual structural response, and to understand the role of the individual beam elasticity in the unit cell response. 

\subsection{Kinematic model}
\label{s:Kin}
\begin{figure}[!htb]
\centering
\includegraphics[scale=1]{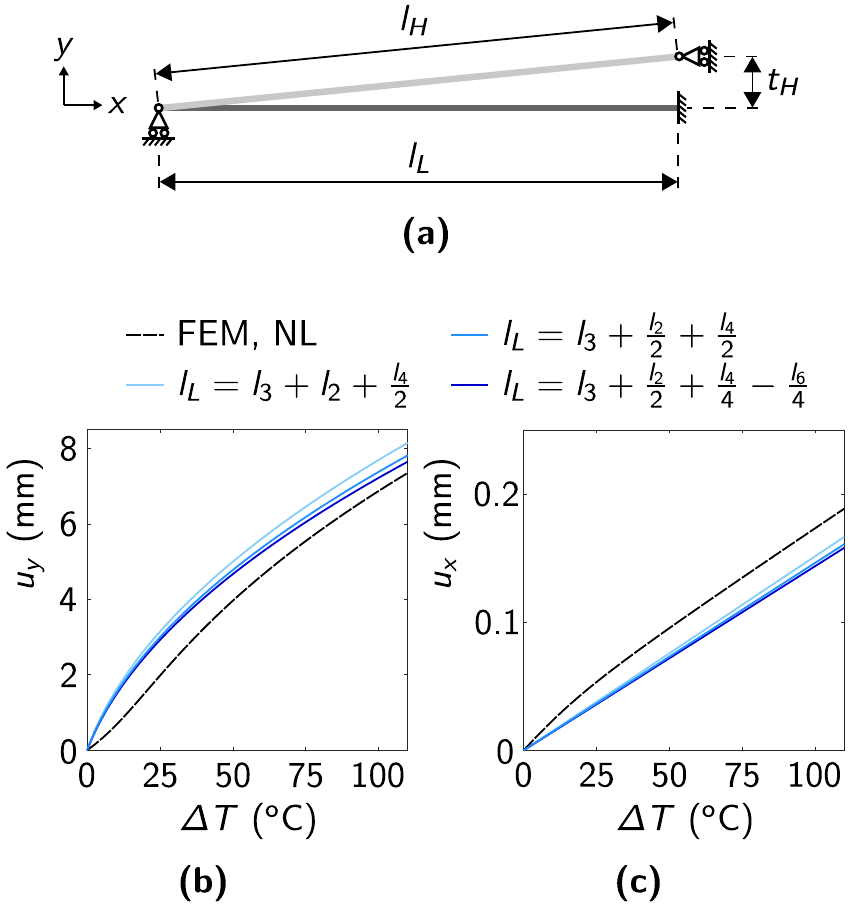}
\caption{(a) Schematic of the pin-jointed representation of a quarter of our unit. (b) Total vertical and (c) horizontal elongation of the unit as a function of the temperature increment. The dashed black line represents the nonlinear FEM reference, while the continuous blue lines come from the kinematic model, for different choices of $l_L$.}
\label{f:Analy}
\end{figure}

As a first step, we analyze a pin-jointed (bar and hinge) analog of our cell. This analog model is illustrated in Fig.~\ref{f:Analy}(a).
Due to symmetry, we consider a quarter of the unit, and we replace it with pin-jointed trusses that can only deform axially. In particular, the low-CTE half-bar is replaced by a bar of length $l_L$. A quarter of the high-CTE frame is replaced by a bar of length $l_H=l_L\cos{\theta}$, where $\theta=\arctan{t_H/l_L}$ is the inclination of the bar in the undeformed configuration and $t_H=t_3-t_2/2-t_4/2$ is the vertical distance between the midpoints of flexure hinges 2 and 4. The choice of $l_L$ strongly affects the response of this purely-kinematic model. We can either choose to account for half of flexure 4 and the whole flexure 2 ($l_L=l_3+l_2+l_4/2$), or half of flexure 4 and half of flexure 2 ($l_L=l_3+l_2/2+l_4/2$), or half of flexure 2 and half of the effective flexure of length $(l_4-l_6)/2$ ($l_L=l_3+l_2+l_4/4$).

Using Gruebler's equation for the kinematics of rigid bodies, we can calculate the number of degrees of freedom (DOFs) of our mechanism:
\begin{equation}
\label{e:dof}
DOF=3(n-1)-2j_1-j_2\,\,,
\end{equation}
where $n=2$ is the number of links (including the ground link) and $j_i$ is the number of joints with $i$ DOFs (e.g., $j_1$ is the number of joints with 1 DOF). In our case, due to the boundary conditions that mimic the symmetry of our unit, we can neglect the horizontal bar and replace the left-most slider with a grounded pin. Thus, we have $n=3$ (the H bar, the vertical slider and the ground link), $j_1=3$ (two pin-joints and one slider), and $j_2=0$. Thus, Eq.~\ref{e:dof} yields $DOF=0$; this highlights that our mechanism reduces to a statically determinate structure and that any motion can be ascribed solely to thermal expansion.

As we increase the temperature from $T_i$ to $T_f=T_i+\Delta T$ and both bars elongate, the low-CTE one remains horizontal due to the boundary conditions stemming from symmetry. Its length after expansion is 
\begin{equation}
\label{e:llowtf}
l_{L}(T_f)=l_{L}\left(1+\alpha_{L}\Delta T\right)\,\,.
\end{equation}
The high-CTE bar expands more than the low-CTE one since we chose $\alpha_H>\alpha_L$; its final length is
\begin{equation}
\label{e:lhightf}
l_{H}(T_f)=l_{H}\left(1+\alpha_{H}\Delta T\right)\,\,,
\end{equation}
The increased expansion is accommodated by a rotation of the high-CTE bar about the left-most pin, which causes the overall structure to elongate vertically. The total vertical displacement of the unit, described by a kinematic model of  a quarter of the structure, is calculated as
\begin{equation}
\label{e:uyupper}
u_y(T_f)=2\left(\sqrt{l_{H}(T_f)^2-l_{L}(T_f)^2}-t_{H}\right)\,\,,
\end{equation}
and its dependency on $\Delta T$ is shown in Fig.~\ref{f:Analy}(b). We can see that this model overestimates the vertical displacement with respect to the nonlinear FE model, regardless of the choice of $l_H$. In particular, this overestimation is more pronounced for small $\Delta T$ values. The case that better resembles our numerics is $l_L=l_3+l_2+l_4/4-l_6/4$, which represents pin joints placed at the middle of the flexure hinges, as commonly done in compliant mechanism analysis ~\cite{howell2001}. 

There are two main reasons why this model does not fully capture the behavior of our structure: i) the kinematics are based on the assumption of pin-joints that, as mentioned, is sensitive to the selection of $l_L$; ii) after thermal expansion, the length of $l_L$ in our model is not affected by the presence of the high-CTE frame. In reality this frame, expanding more than the inner bar, pulls the bar outwardly causing it to elongate further. This is visible in Fig.~\ref{f:Analy}(c), where we plot the elongation of the low-CTE bar and  compare numerical and theoretical  predictions; we can clearly see that the latter  underestimates the inner bar elongation. This underestimation is worse if we select $l_L$ to be shorter than its actual value $l_L=l_3+l_2+l_4/2$.

\subsection{Mechanistic model}
\label{s:Mech}
\begin{figure}[!htb]
\centering
\includegraphics[scale=1]{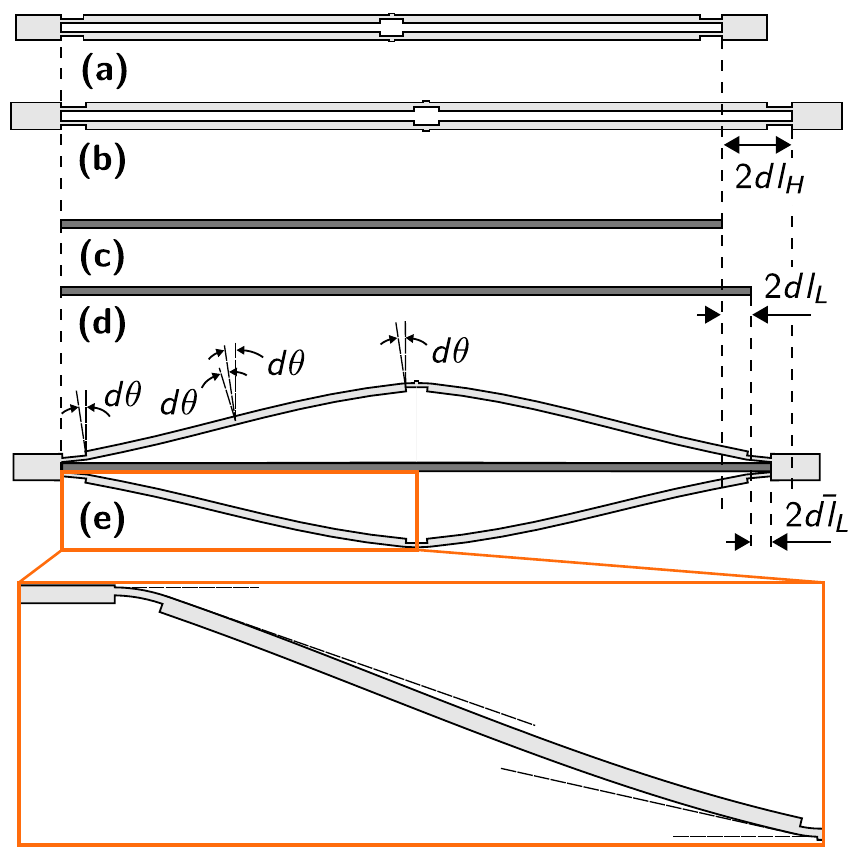}
\caption{(a-e) Schematics illustrating that, due to thermal expansion, the low-CTE bar elongates more than its nominal extension, i.e., it elongates of $2dl_L+2\bar{dl_L}$ instead of $2dl_L=\alpha_L2l_L(\Delta T)$. (a,b) Standalone high-CTE frame before and after thermal expansion. (c,d) Standalone low-CTE bar before and after thermal expansion. (e) Magnified deformation of the whole unit after thermal expansion{, with a detail showing how the beams are deforming (the dashed lines, parallel to the beams at their attachment points, should help visualize bending deformations).}}
\label{f:MechUnit}
\end{figure}

Given the limitations of the purely-kinematic model, we develop a model that accounts for the elasticity of the high- and low-CTE portions of the beams. To understand its main features, refer to the schematics of Fig.~\ref{f:MechUnit}. Note that these images are manipulated FEM results where the displacements are magnified to aid our explanation. 
Considering the outer frame as a standalone element, thermal expansion causes its inner void to elongate of an amount $2dl_H=2l_L \alpha_H \Delta T$, as shown in Fig.~\ref{f:MechUnit}(a,b), where $l_L$ is chosen as $l_1/2$ to be consistent with Fig.~\ref{f:Analy}(a). Considering the inner bar as a standalone element, thermal expansion causes it to elongate of $2dl_L=2l_L \alpha_L \Delta T$, as shown in Fig.~\ref{f:MechUnit}(c,d). Since $\alpha_H>\alpha_L$, we have that $dl_H>dl_L$. Thus, when we consider the two parts together as shown in Fig.~\ref{f:MechUnit}(e), the outer frame will pull outwardly on the low-CTE bar, causing it to elongate of an additional amount $2\bar{dl_L}$.

As a first step in our model, we derive the potential energy of the high-CTE frame, and use it to determine the force exerted by this part on the low-CTE bar. Our first assumption, motivated by the magnified FEM result shown in Fig.~\ref{f:MechUnit}(e), is that all beams of the outer frame deform by pure bending. We further assume that, after deformation, the end angle of beams of type 2 is $d\theta$, the end angle of the flexures having length $l_4/2-l_6/2$ is $d\theta$, and each half of the type 3 beams is bent of $d\theta$. Recalling that the potential energy of a beam in pure bending with an end angle $\theta$ can be calculated as
\begin{equation}
U=\frac{EI}{2l}\theta^2\,\,,
\end{equation}
the total potential energy of our structure is
\begin{equation}
U_{T}=2\frac{E_HI_2}{l_2}d\theta^2+4\frac{E_HI_4}{l_4-l_6}d\theta^2+8\frac{E_HI_3}{l_3}d\theta^2\,\,,
\label{e:utot}
\end{equation}
where $I_2$, $I_3$, $I_4$ are the second moments of area of beams of type 2, 3 and 4, respectively. Assuming small angles, an assumption that is reasonable when the cell is much longer along $x$ than along $y$, and considering small temperature increments (see \ref{a:theta}), we can rewrite $d\theta$ as
\begin{equation}
d\theta=\frac{dl_H-dl_L-\bar{dl_L}}{t_H+u_y/2}\,\,,
\end{equation}
were $dl_H$, $dl_L$ and $\bar{dl_L}$ have been previously defined and are illustrated in Fig.~\ref{f:MechUnit}(b,d), $u_y$ is here intended as the overall vertical displacement of the structure at the previous temperature increment and $t_H=t_3-t_2/2-t_4/2$. In light of this definition, we can rewrite the potential energy as
\begin{multline}
U_{T}=\frac{1}{2}\left(4\frac{E_HI_2}{l_2(2t_H+u_y)^2}+8\frac{E_HI_4}{(l_4-l_6)(2t_H+u_y)^2}\right.\\
\left. +16\frac{E_HI_3}{l_3(2t_H+u_y)^2}\right)(2dl_H-2dl_L-2\bar{dl_L})^2\\
=\frac{1}{2}k_H(2dl_H-2dl_L-2\bar{dl_L})^2\,\,.
\label{e:Utot}
\end{multline}
In the last equation, we defined $k_H$ as the axial stiffness of the high-CTE frame. Taking a derivative with respect to $(2dl_H-2dl_L-2\bar{dl_L})$, we obtain an expression for the axial force exerted by the high-CTE frame on the low-CTE bar:
\begin{equation}
F_H=k_H(2dl_H-2dl_L-2\bar{dl_L})\,\,.
\end{equation}

\begin{figure}[!htb]
\centering
\includegraphics[scale=1]{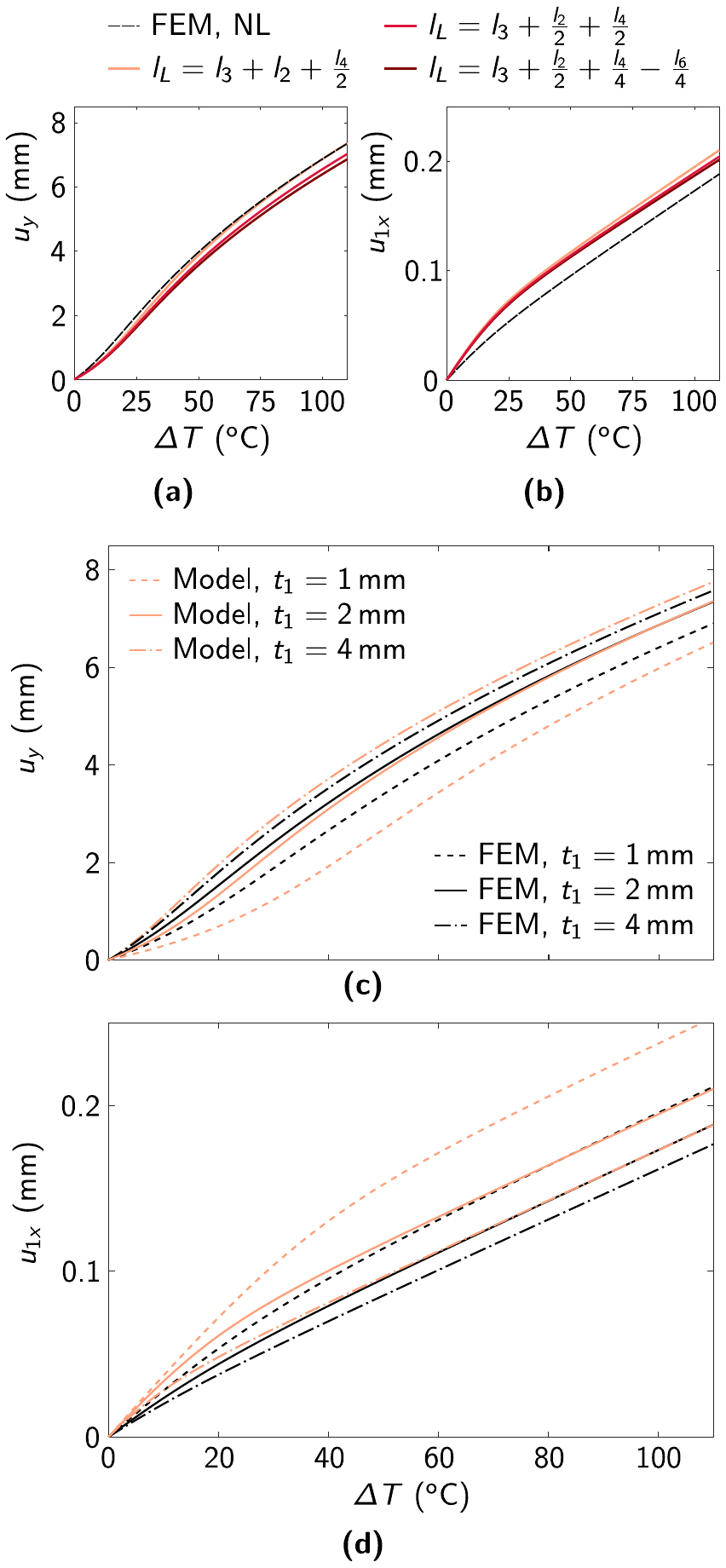}
\caption{(a) Total vertical elongation of the unit and (b) elongation of the low-CTE bar as a function of the temperature increment. The dashed black line represents the nonlinear FEM reference, while the continuous red lines are obtained from the mechanistic model, for different choices of $l_L$. (c) Total vertical elongation of the unit and (d) elongation of the low-CTE bar for different values of $t_1$ (thickness of the low-CTE bar), for $l_L=l_3+l_2+l_4/2$.}
\label{f:MechResults}
\end{figure}

With  the axial stiffness of the high-CTE frame determined, we can now solve the force equilibrium at the interface between the two materials and derive the updated elongation of the low-CTE bar, $dl_L+\bar{dl_L}$. To do so, we set
\begin{equation}
F_H=k_H(2dl_H-2dl_L-2\bar{dl_L})=k_L2\bar{dl_L}=F_L\,\,,
\end{equation}
where $k_L=E_LA_1/(2l_L)$ is the axial stiffness of the low-CTE bar. Manipulating this last equation yields an expression for the updated elongation:
\begin{equation}
\bar{dl_L}=\frac{k_H}{k_H+k_L}(dl_H-dl_L)\,\,.
\label{e:bal}
\end{equation}
This calculation needs to be repeated at each temperature increment; at each step, we update $dl_H$, $dl_L$, but also $k_H$, recalling that the latter is a function of $u_y$. In our computations, where we consider a discrete number of small temperature increments ($<1\,\mathrm{\degree C}$), we consider $u_y=u_y(\Delta T_{n-1})$ at increment $\Delta T_n$. 

Finally, we use the updated half-bar elongation $dl_L+\bar{dl_L}$ to derive the vertical displacement for the current temperature increment. This step is carried out using the kinematic description derived in the previous section and sketched in Fig.~\ref{f:Analy}(a). First of all, we calculate the extensions of the high- and low-CTE bars as
\begin{equation}
l_L(T_f)=l_L(T_i)+dl_L+\bar{dl_L}\,\,,
\end{equation}
\begin{equation}
l_{H}(T_f)=l_{H}\left(1+\alpha_{H}\Delta T\right)\,\,.
\end{equation}
Then, we insert  these values into Eq.~\ref{e:uyupper} to find the overall vertical displacement of our structure, $u_y(T_f)$. Please note that, once again, the choice of $l_L$ (and, consequently, of $l_H=\sqrt{t_H^2+l_L^2}$) affects the results as  illustrated in Fig.~\ref{f:MechResults}(a,b). Here the vertical displacement of the whole unit and the horizontal displacement of the low-CTE bar ($u_{1x}$) are plotted  as a function of the temperature increment for given values of $l_L$. The curves are compared to the those obtained from the nonlinear FE model. For all values of $l_L$, we can see that the mechanistic model captures the mechanics of the structure more accurately than the purely-kinematic one. This is particularly clear from the $u_y$ plot, where we can appreciate that the curves from the model capture the trends of the numerical results, i.e., the convexity near the origin that morphs into a concavity as the temperature increases. From the $u_{1x}$ plot, we can observe  that the model generally overestimates the horizontal elongation of the low-CTE bar; however, the prediction is still superior with respect to the purely-kinematic one in Fig.~\ref{f:Analy}(c) and Section~\ref{s:Kin}, which did not account for any influence of the high-CTE frame on the low-CTE elongation. If we choose $l_L$ to include the whole flexure 2 and half of flexure 4 ($l_L=l_3+l_2+l_4/2$), the vertical elongation matches the numerics very well. For this reason, we consider this value of $l_L$ in the remainder of the article. 

To probe the accuracy of the model, we report results for three values of $t_1$, the thickness of the low-CTE bar. Recall that the value we use up to this point is $t_1=2\,\mathrm{mm}$. Fig.~\ref{f:MechResults}(c) shows that $t_1=2\,\mathrm{mm}$ is the value for which the model best matches the FE results. If we decrease $t_1$ to $1\,\mathrm{mm}$ or if we increase it to $4\,\mathrm{mm}$, thereby respectively increasing or decreasing the total longitudinal elongation of the cell after thermal expansion with respect to the $t_1=2\,\mathrm{mm}$ case, the discrepancy in both $u_{1x}$ and $u_y$ increases. We ascribe these discrepancies to two main reasons. First, the assumption that each half of the type 3 beams bends of an angle $d\theta$ is not always exact since bulky flexures might require to keep into account shear effects. Second, using pin-jointed kinematics to predict the total vertical elongation of the cell can also lead to inaccuracies.

\begin{figure}[!htb]
\centering
\includegraphics[scale=1]{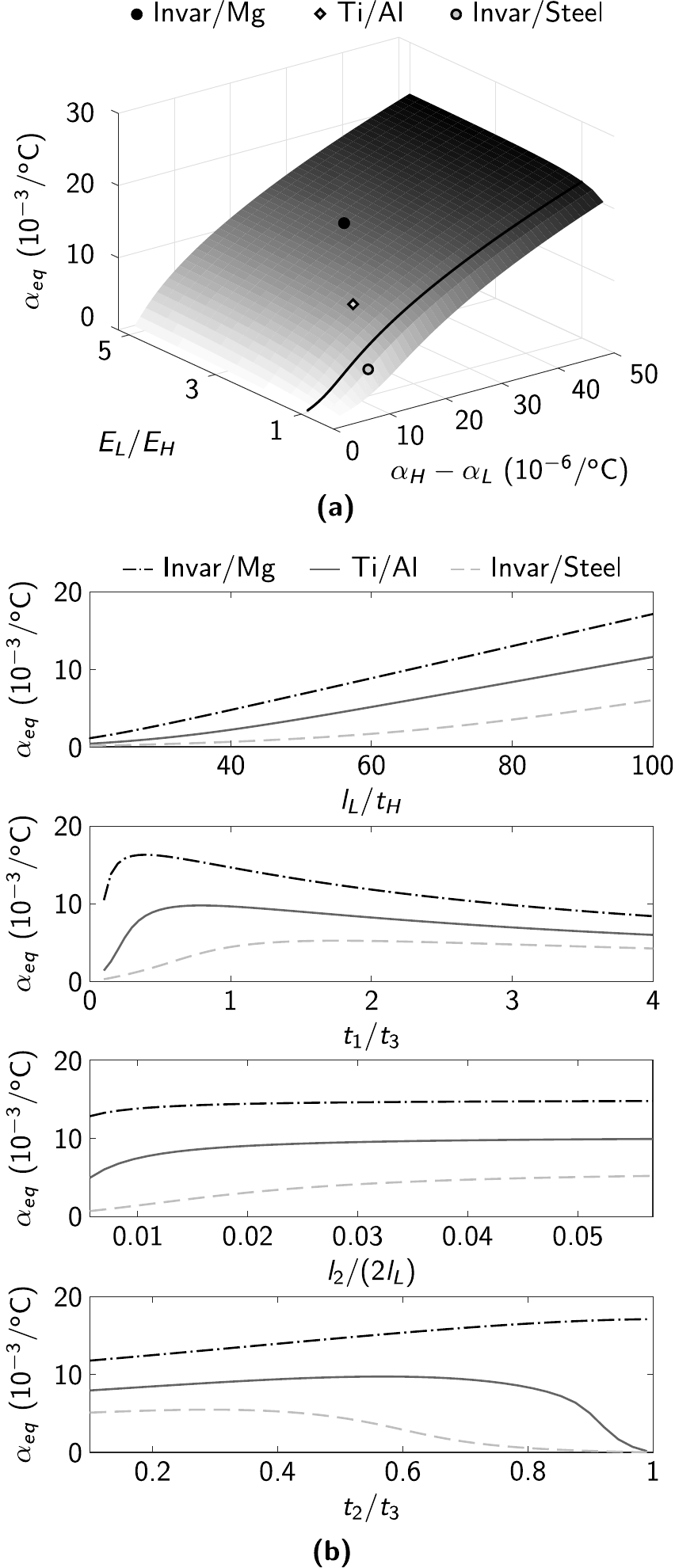}
\caption{Mechanistic prediction of the influence of the design parameters on the equivalent CTE of one unit $\alpha_{eq}$. (a) Influence of the material properties (Young's moduli and CTE mismatch) for the default geometry parameters of Sec.~\ref{s:Num}. The thick black line marks $E_L/E_H=1$. (b) Influence of the geometrical parameters, for three material couples. From top to bottom, we analyze: the ratio between $t_H=t_3-t_2/2-t_4/2$ and the low-CTE bar half-length $l_L$; the thickness of the low-CTE bar, $t_1$ (with $t_3$ constant); the flexure length $l_2=l_4$ (with $l_1=2l_L$ constant); the flexure thickness ratio $t_2/t_3$.}
\label{f:Param}
\end{figure}

\subsection{Influence of the design parameters}
The mechanistic model can now help us assess the influence of the design parameters on the overall thermal expansion of our structure. To do so, we plot the equivalent thermal expansion ($\alpha_{eq}$) of the unit in the direction of maximum displacement, i.e., along $y$. First, we set the geometric parameters to the default values introduced in Sec.~\ref{s:Num}, and analyze the influence of the material parameters, i.e., the Young's moduli ratio $E_L/E_H$ and the mismatch between CTEs of the base materials ($\Delta\alpha = \alpha_H-\alpha_L$). Fig.~\ref{f:Param}(a) shows the design map, where the circular markers highlight the results for three material couples of interest: (i) Ti as low-CTE and Al as high-CTE metal; (ii) Invar as low-CTE and Mg (alloy AZ31B) as high-CTE metal, where we select $E_L=141\,\mathrm{GPa}$, $\alpha_L=1.6\,\,\mathrm{10^{-6}/\degree C}$, $E_H=45\,\mathrm{GPa}$, $\alpha_H=26\,\,\mathrm{10^{-6}/\degree C}$; (iii) Invar as low-CTE and Steel (AISI 1040) as high-CTE metal, where we select $E_L=141\,\mathrm{GPa}$, $\alpha_L=1.6\,\,\mathrm{10^{-6}/\degree C}$, $E_H=193\,\mathrm{GPa}$, $\alpha_H=11.3\,\,\mathrm{10^{-6}/\degree C}$. We can observe that, predictably, increasing $\Delta\alpha$ causes a significant increase of $\alpha_{eq}$. The influence of $E_L/E_H$ is minimal unless we select values of $\alpha_{eq}$ close to or smaller than one, e.g., in the case of Invar-Steel. The surface plot highlights that the $\alpha_{eq}$ of the unit structure is three orders of magnitude larger than the CTE of the constituent materials; in particular $\alpha_{eq}\approx9.6\,\,\mathrm{10^{-3}/\degree C}$ for Ti/Al, $\alpha_{eq}\approx14.7\,\,\mathrm{10^{-3}/\degree C}$ for Invar/Mg, and $\alpha_{eq}\approx4.6\,\,\mathrm{10^{-3}/\degree C}$ for Invar/Steel.   

For the three material couples mentioned above, we now assess the influence of the geometrical parameters of the unit. We show all these results in Fig.~\ref{f:Param}(b), where $\alpha_{eq}$ is plotted versus four nondimensional parameters: $l_L/t_H$, $t_1/t_2$, $l_2/(2l_L)$ and $t_2/t_3$. Note that we set $l_2=l_4$ and $t_2=t_4$, and that all the geometrical parameters not being analyzed are set to the default values of Section~\ref{s:Num}. The first plot from the top in Fig.~\ref{f:Param}(b) shows the influence of the aspect ratio of the unit, $l_L/t_H$ ($l_L$ is the half-length of the low-CTE bar, and $t_H=t_3-t_2/2-t_4/2$ is the mismatch in height between the centerpoints of flexures 4 and 2). 
We observe that an increase of $l_L/t_H$ causes an increase of $\alpha_{eq}$. 
To gauge the influence of the elastic moduli of the high- and low-CTE parts, we plot the equivalent thermal expansion as a function of the ratio $t_1/t_3$, knowing that this parameter affects the effective stiffness of our structure. Increasing $t_1$ causes the structure to be stiffer axially and to elongate more along $y$, as also shown in Fig.~\ref{f:MechResults}(d) and as discussed in Sec.~\ref{s:Mech}. However, larger $t_1$ values imply that a certain vertical displacement yields a smaller vertical strain. The latter effect is clearly dominating, as $\alpha_{eq}$ decreases with increasing $t_1/t_3$. Below $t_1/t_3=0.1$, the low-CTE bar has a vanishing axial stiffness; this, in turn, causes the high-CTE frame to become insensitive to the axial stiffness of the low-CTE bar and $\alpha_{eq}$ to drop. Changing the parameter $l_2/(2l_L)$, i.e., the length of the flexure divided by the overall length of the low-CTE bar, has minor effects on $\alpha_{eq}$ as we show in the third plot from the top in Fig.~\ref{f:Param}(b). Note that too small values of $l_2/(2l_L)$ are bound to break the assumption regarding the angles subtended by the various beams during deformation. This assumption is also broken for large values of $t_2/t_3$, i.e., when the thickness of the flexure hinge is comparable to the thickness of the thicker high-CTE beams (beams of type 3). From the $t_2/t_3$ plot, we can see that the three material couples behave differently: in the neighborhood of the default value $t_2/t_3=0.5$, $\alpha_{eq}$ increases with $t_2/t_3$ for Invar/Mg, keeps constant for Ti/Al and decreases for Invar/Steel. This sensitivity to the choice of material stems from the fact that the influence of the Young's modulus and that of the flexure thickness $t_2$ are intertwined and affect the balance between $k_H$ and $k_L$ when determining the longitudinal elongation of the unit (see Eq.~\ref{e:bal}). In addition, since the stress localizes at these flexures during deformation, we can further appreciate how the flexure design is critical for these structures.

\section{Experimental validation}
\label{s:Exp}
\begin{figure}[!htb]
\centering
\includegraphics[scale=1]{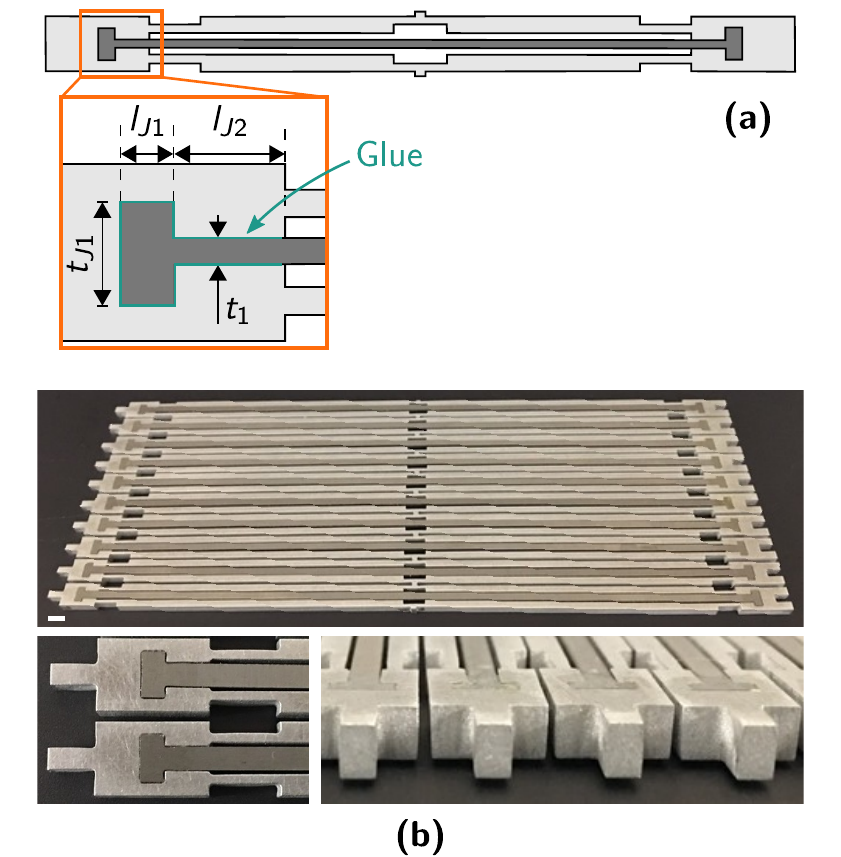}
\caption{(a) Schematic of the modified structure, featuring jigsaw -like joints designed to provide tight connections between the metals. In addition, a bonding agent (Loctite E-120 HP Hysol) was deposited at the interface of the metals (green) to further improve their bonding. (b) Picture of one specimen, with details highlighting the quality of the water jet cutting and the presence of some out-of-plane tapering. Please note that male/female connectors are added at the left and right edges of the arrays, to facilitate their assembly into three-dimensional structures. Scale bar: 0.5 cm.}
\label{f:ExpSpec}
\end{figure}

To validate our numerical and theoretical predictions, we manufacture and test a set of bi-metal specimens. We use water-jet cutting (OMAX MicroMAX, that produces cuts of $0.3\,\mathrm{mm}$ width) on $3\,\mathrm{mm}$-thick sheets of aluminum (Al-6061-T6) and titanium (Ti-6-4), and we manually insert the Ti bars into the Al frame. The specimens are linear arrays of units with dimensions identical to those of our numerical models (with $t_1=2\,\mathrm{mm}$; refer to Sec.~\ref{s:Num} for details). To ensure a sturdy fixation of the two metals joining at both ends, we conceived a jigsaw-like joint, inspired by the work of Steeves et al.\ on low-CTE lattices~\cite{steeves2007}, where the two metals interlock with a tight fit, as shown in Fig.~\ref{f:ExpSpec}(a).
\begin{figure}[!htb]
\centering
\includegraphics[scale=1]{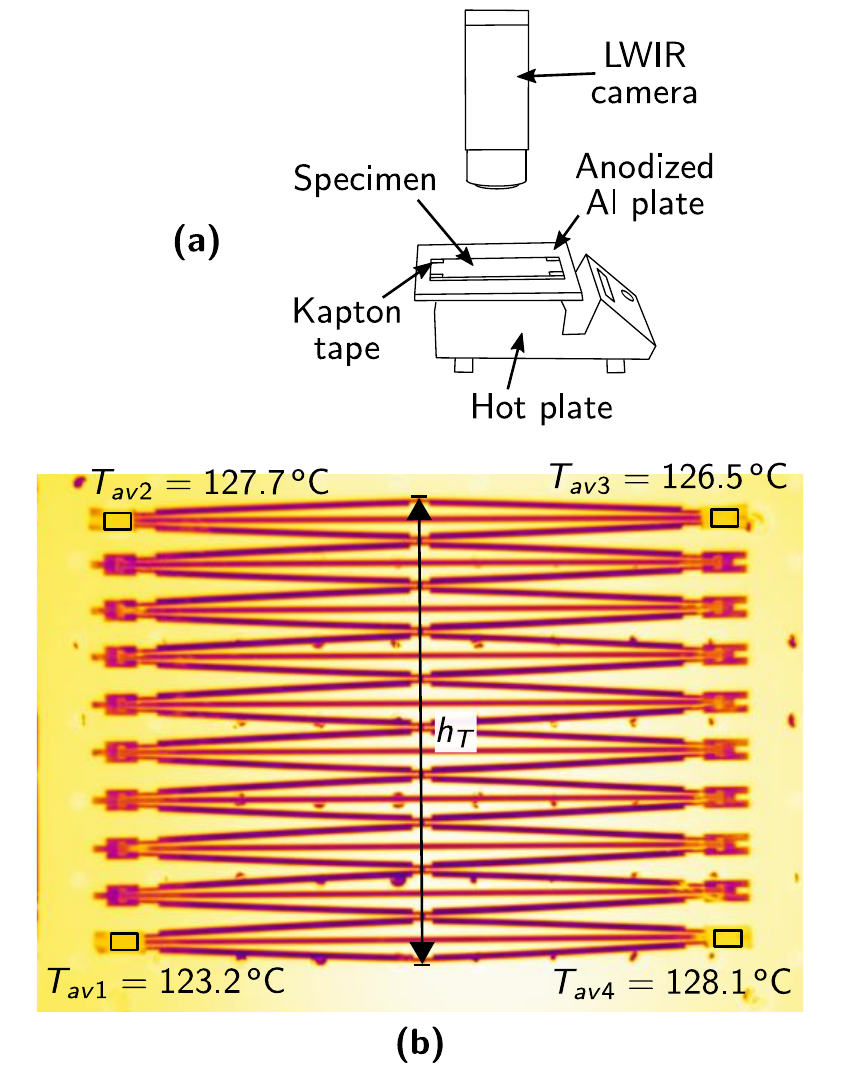}
\caption{(a) Experimental setup. (b) Representative image acquired from the thermal camera. The black rectangles indicate areas where the Kapton tape was placed, and where the average temperature was measured. The temperature of the specimen is then obtained as  $T_{av}=(T_{av1}+T_{av2}+T_{av3}+T_{av4})/4$. In this specific case, the total height of the specimen at $T_{av}=126.4\,\mathrm{^oC}$ is $h_T\approx146\,\mathrm{mm}$.}
\label{f:ExpSetup}
\end{figure}
In particular, we select parameters $t_{J1}=4\,\mathrm{mm}$, $l_{J1}=2\,\mathrm{mm}$ and $l_{J2}=4\,\mathrm{mm}$ for the joints. One of our specimens, an array of 10 units, is shown in Fig.~\ref{f:ExpSpec}(b). The detail to the left shows the top view of one of the junctions. 
To facilitate the assembly of multiple specimens, we also add male/female connectors at the left and right edges of each unit, respectively, as also shown in Fig.~\ref{f:ExpSpec}(b). From these photos, we can appreciate the precision of the water jet cutting process that, unlike conventional water jet cutting, does not cause unwanted material removal at sharp corners. The detail on the right shows that water jetting produces some out-of-plane tapering of the specimen. We manufacture three of these 10 unit arrays; in all three cases, we bonded high- and low-CTE parts as illustrated in Fig.~\ref{f:ExpSpec}(a).
Three types of bonding strategies are examined. Specimen 1 is bonded by applying super glue (cyanoacrylate) on the connectors before assembling the various parts. Specimen 2 is first assembled and then bonded with epoxy (Loctite E-120 HP Hysol{, that can withstand temperatures up to $150\,\mathrm{^oC}$}) applied on the surface of the structure. Specimen 3 is bonded by applying the same epoxy glue on the connectors before assembling them. As shown later, the bonding process affects the quality of the connections and hence the specimen response.

A sketch of our experimental setup is shown in Fig.~\ref{f:ExpSetup}(a). The specimens are placed on an Anodized Aluminum plate, which is directly placed on a hot plate (Thermo Scientific Super-Nuova). The temperature and deformation of the specimen are measured via a Longwave Infrared camera (FLIR A655sc). To monitor the specimen temperature, and to avoid reflections from the metallic surfaces, we cover parts of the specimen with non-reflective Kapton tape. Our measurement consists of multiple steps. First, we heat up the anodized aluminum plate at a desired temperature, that is monitored directly with the thermal camera. 
We place our specimen on the aluminum plate, and we measure the temperature on the Kapton tape patches placed at the four corners. Fig.~\ref{f:ExpSetup}(b) shows an example of an image acquired from the thermal camera. The average temperature was measured in each region within the black rectangles, i.e., $T_{av1}$, $T_{av2}$, $T_{av3}$ and $T_{av4}$. When the difference between these averages is  below $5\,\mathrm{\degree C}$, we acquire an image and assign it the average temperature $T_{av}=(T_{av1}+T_{av2}+T_{av3}+T_{av4})/4$, and we measure the overall specimen height $h_T$. 
From $h_T$, we derive the overall specimen elongation; dividing it by the number of cells, we obtain the single unit $u_y$ at several temperature increments, and we compare the results with the numerical and theoretical predictions. To account for measurement errors, which can be attributed to frictional effects, dimensional inaccuracy in measuring lengths from pixelated images, and inhomogeneous heating, we repeat each experiment three times (this is the bare minimum for repeatability). Thus, each experimental point indicates the mean of three measured values, and the error bar represents one standard deviation. 

\begin{figure}[!htb]
\centering
\includegraphics[scale=1]{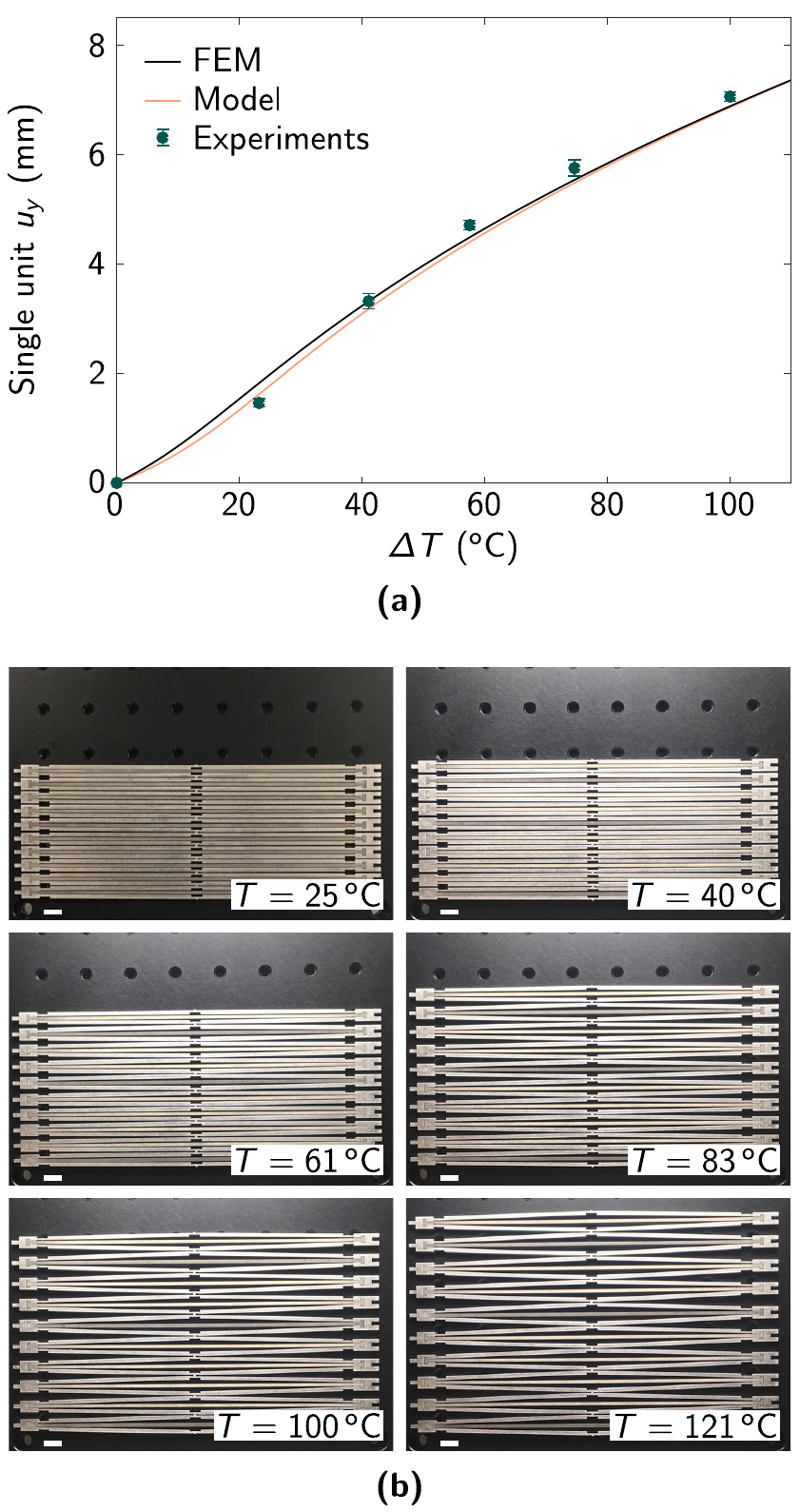}
\caption{(a) Comparison between numerical, theoretical and experimental values representing. the vertical displacement of a single unit of our 10-unit specimen. (b) Deformed specimens at given temperature values. Scale bar: 1 cm.}
\label{f:ExpArray}
\end{figure}

Fig.~\ref{f:ExpArray}(a) shows the comparison between the experimental results and those from theory and computations for specimen 3, where epoxy was applied to bond the metals before assembly.
\begin{figure}[!htb]
\centering
\includegraphics[scale=1]{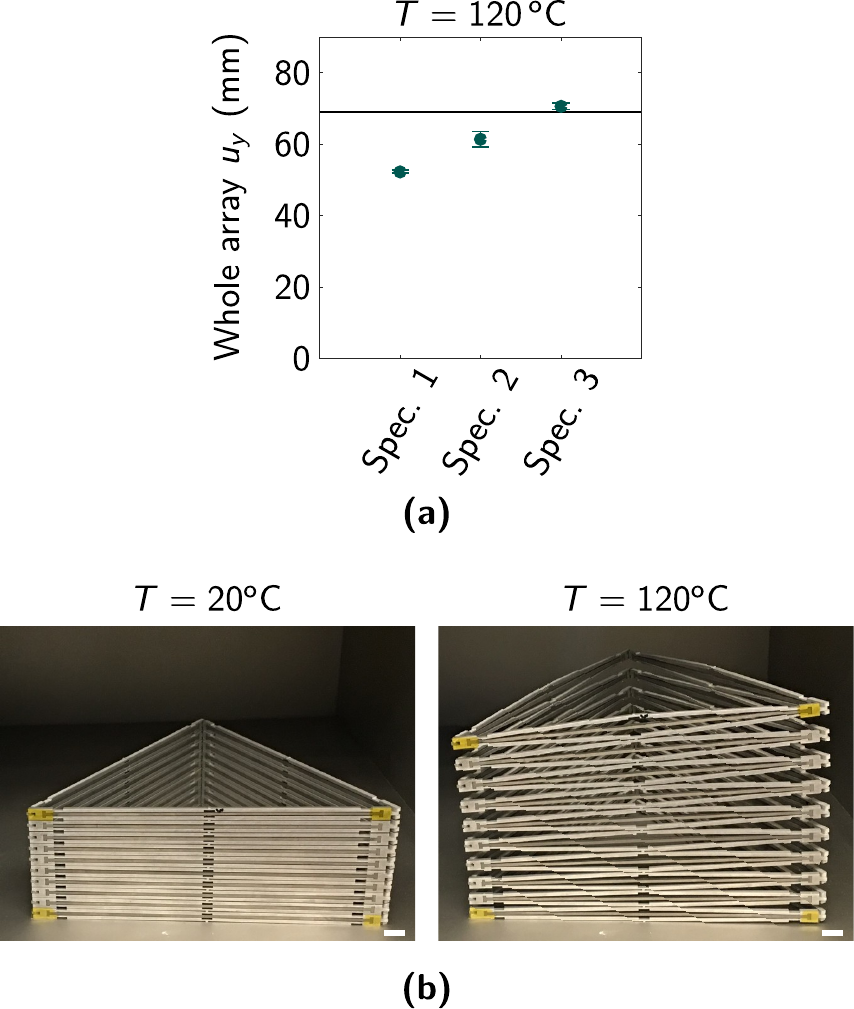}
\caption{(a) Comparison between the whole array elongation $u_y$, for specimens 1, 2 and 3 at $120\,\mathrm{\degree C}$. Specimen 1 is bonded with a super (cyanoacrylate) glue. Specimen 2 is bonded by applying epoxy after the manual assembly of various parts. Specimen 3 is obtained by applying epoxy at the interface between metals at the junctions before assembly. The horizontal line marks the expansion value from a numerical simulation assuming perfect bonding. (b) Demonstration of a three-dimensional thermally-expanding structure with a triangular footprint, obtained by assembling the three specimens discussed in (a) by means of the male/female connectors shown in Fig.~\ref{f:ExpSpec}(b). Scale bar: 1 cm.}
\label{f:Exp3D}
\end{figure}
For this specimen, the results from experiments, computations and theory are in excellent agreement. Fig.~\ref{f:ExpArray}(b) shows the temperature-induced deformation with snapshots illustrating the deformed specimen at various temperatures. 
To analyze the effects of the bonding agent (cyanoacrylate versus epoxy) and bonding strategy (application before versus after the assembly) on the deformation, the experiment is repeated for specimens 1 and 2. 
Fig.~\ref{f:Exp3D}(a) shows the comparison between the total displacement (of all 10 units) of the three specimens at $T=120\,\mathrm{\degree C}$. The results differ considerably, thereby highlighting the strong influence of the bonding agent and of the temporal sequence of the application steps. In particular, application of the epoxy prior to assembly results in the largest overall displacement, a result aligned (albeit slightly higher) with  the numerical prediction for the same array. On the other hand, applying epoxy on the surface of a specimen assembled via frictional forces produces an imperfect bonding and a lower overall expansion with respect to the perfectly bonded simulated results. Finally, cyanoacrylate provides the worse bonding with a resulting overall expansion that is 25\% lower than the numerical prediction. 

As a final experiment, we use the same three array specimens described in Fig.~\ref{f:Exp3D}(a), each with units assembled with a specific bonding strategy, to generate a three-dimensional structure with a triangular footprint, i.e., a triangular prism (Fig.~\ref{f:Exp3D}(b)). The connections between the three arrays relies solely on friction and no bonding agent is used. The specimen is placed in a heating chamber and heated up to $120\,\mathrm{\degree C}$. From the right panel of Fig.~\ref{f:Exp3D}(b), we can see that the specimen expands considerably. While this setup does not allow to accurately quantify the deformation, we can make the following qualitative remarks: The vertical expansion is not uniform, i.e. the triangular faces of the prism do not remain parallel during expansion, as it would be for an assembly of arrays with identical geometry and bonding strategy. In contrast, the specimen tilts slightly on one side, because each array of the triangular layout uses a distinct bonding strategy that leads to a specific rate of expansion, as shown in Fig.~\ref{f:Exp3D}(a). This experiment demonstrates that the assembly of arrays expanding uniaxially at different rate can be exploited to generate deformation other than uniaxial, hence opening up venues to explore other modes of out-of-plane deformation.

\begin{figure}[!htb]
\centering
\includegraphics[scale=1]{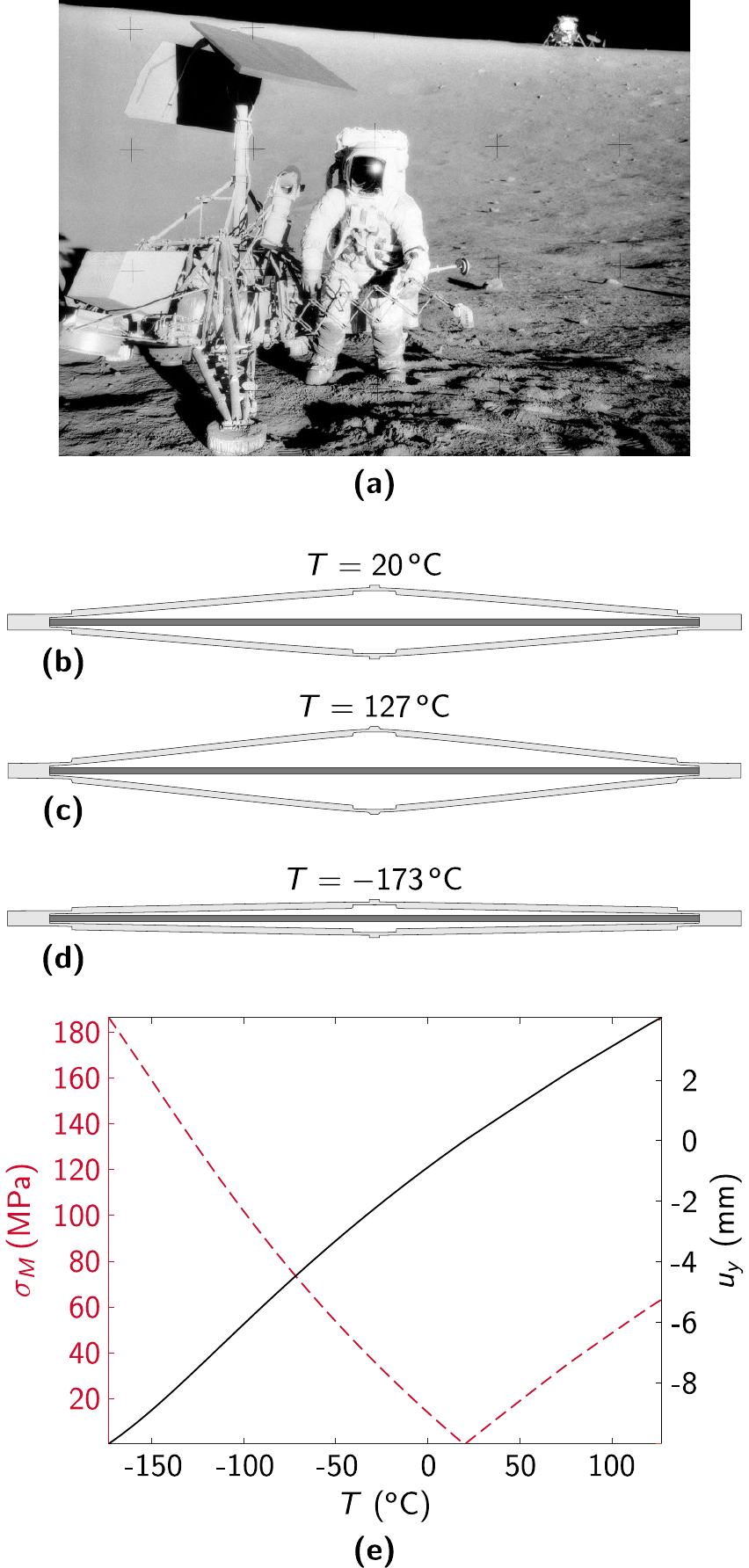}
\caption{(a) Photo of the surveyor III lander (source: NASA). (b) Undeformed structure designed to expand over the typical temperature range of the surface of the moon (-$173\,\mathrm{\degree C}$ to $127\,\mathrm{\degree C}$). (c,d) Deformed structure at the maximum daily temperature and the minimum nightly temperature, respectively. (e) This plot highlights both the overall vertical displacement of the unit, $u_y$ and the maximum von Mises stress recorded during deformation $\sigma_M$. Please note that this stress is recorded at the shortest flexure, as discussed in Sec.~\ref{s:Num}, and that it remains below the Yield stress of aluminum $\sigma_Y=260\,\mathrm{MPa}$.}
\label{f:Lunar}
\end{figure}

\section{Behavior over lunar temperature ranges}
\label{s:lun}
The results shown in the previous sections for temperatures ranging from $20\,\mathrm{\degree C}$ to $130\,\mathrm{\degree C}$ highlight that it is possible to design bi-metallic structures capable of achieving large effective CTEs. However, special considerations have to be made when they need to withstand both positive and negative temperatures. This scenario becomes relevant in the context of lunar structures, since the typical temperature on the surface of the moon ranges from -$173\,\mathrm{\degree C}$ to $127\,\mathrm{\degree C}$. A famous example of lunar spacecraft is the Surveyor III lander, shown in Fig.~\ref{f:Lunar}(a) (image from NASA). As many space-faring agencies are looking into both human and robotic lunar landers, structures that harness these predictable day-night temperature shifts to perform functions are becoming increasingly relevant, since they would allow the spacecraft to survive the lunar night or would allow the spacecraft to deploy a component at the day/night interface using only the temperature change. Some examples of how these structures could be used in the lunar environment include passive switches~\cite{heo2019} that connect a solar array to a battery during the day, while removing the connection at night. This would be designed to save the battery power to enable the survival of a non-nuclear-powered rover during the lunar night. They could also be used for deploying or retracting an antenna, radiator or solar panel to either activate or deactivate them during the day/night transition. We also envision that they could be used as a dust cover for a lunar telescope that only needs to open during the night, but it then needs to be closed during the day to protect the mirror from dust accumulation. Here, we limit ourselves to the preliminary task of using numerical models to design a structure that expands without plastic deformations over the lunar temperature range. In particular, for our design, we once again select Al-6061-T6, whose yield stress is $\sigma_Y=260\,\mathrm{MPa}$. To provide realistic results that account for the temperature dependence of the linear coefficient of thermal expansion $\alpha$, we consider the temperature-dependent CTE of the base materials discussed in~\ref{a:cte} and illustrated in Fig.~\ref{f:CteFull}.

The design needs to account for the fact that the structure, fabricated in a stress-free state at $20\,\mathrm{\degree C}$, has to be able to expand during the lunar day and contract at night, without developing stresses that would result into plastic deformation. To achieve this, we design the structure shown in Fig.~\ref{f:Lunar}(b) and we consider this configuration as our stress-free state at $20\,\mathrm{\degree C}$. Note that, to avoid stresses that exceed the plastic limit of Al-6061-T6, we use a structure featuring most of the geometrical parameters discussed in Section~\ref{s:Num}; the only exception is the length of flexure 4, the zone that featured the largest stresses in Fig.~\ref{f:NumericalResults}, whose length is here doubled to $l_4=2l_2$. As the temperature increases to $127\,\mathrm{\degree C}$, the structure behaves as described in the previous sections: the low-CTE bar constrains the longitudinal expansion of the high-CTE frame, which expands along the vertical direction, as shown in Fig.~\ref{f:Lunar}(c). For decreasing temperature below $20\,\mathrm{\degree C}$, the high-CTE frame shrinks more than the low-CTE bar. Hence, the structure condenses   until it reaches an almost flat state as shown in Fig.~\ref{f:Lunar}(d). The evolution of the overall vertical displacement of the structure over the temperature range of interest is illustrated in Fig.~\ref{f:Lunar}(e). This figure also illustrates that both expansion and contraction cause the flexures to experience a state of maximum von Mises stress below the yield value of $\sigma_Y=260\,\mathrm{MPa}$, hence ruling out the emergence of any plasticity in our structures.  {Note that the adhesives we used in our experiments are not suitable for the lunar temperature ranges. For this application, fabricating the whole structure via multi-metal additive manufacturing would be advantageous, as it would allow us to bypass the use of a bonding agent.}

\section{Conclusions}
\label{s:Concl}
In this article, we demonstrate that it is possible to realize bi-metallic structures capable of undergoing large deformations in response to temperature changes. We present a design featuring bulky beams connected by flexures and showed how its expansion in response to temperature variations is governed by the constituent material and the choice of geometrical parameters. We propose a theoretical model, validated by experiments, that takes into account both the kinematics and the elasticity of the structure. We use the model to investigate the design space of the structure, varying its geometry and constitutive materials properties. We also use our numerical models to design a structure that can expand without plastically deforming, over the temperature ranges typical of the lunar day-night cycle. Our work lays the groundwork for the development of spacecraft structures that respond to temperature swings, without having to rely on motors and gearing. This could be useful for reducing parts count and increasing reliability of deployable structures.


\section*{Acknowledgment}
This research was carried out at the California Institute of Technology and the Jet Propulsion Laboratory under a contract with the National Aeronautics and Space Administration, and funded through the President's and Director's Fund Program. S.T. acknowledges the support of Roketsan. P.C. and C.D. acknowledge support from the Foster and Coco Stanback Space Innovation Fund. The authors wish to thank Fernando Garza for his fabrication efforts, Christina Naify and Terry Gdoutos for useful discussions.

\appendix
\section{Deformed angle ($d\theta$) derivation}
\label{a:theta}
This appendix illustrates the derivation of the angle $d\theta$, i.e., the end angle of the flexures of type 2 and of the flexures having length $l_4/2-l_6/2$ after deformation. This angle is used in Section~\ref{s:Mech} to obtain a formula for the total potential energy as a function of $dl_H-dl_L-\bar{dl_L}$. 

\begin{figure}[!htb]
\centering
\includegraphics[scale=1]{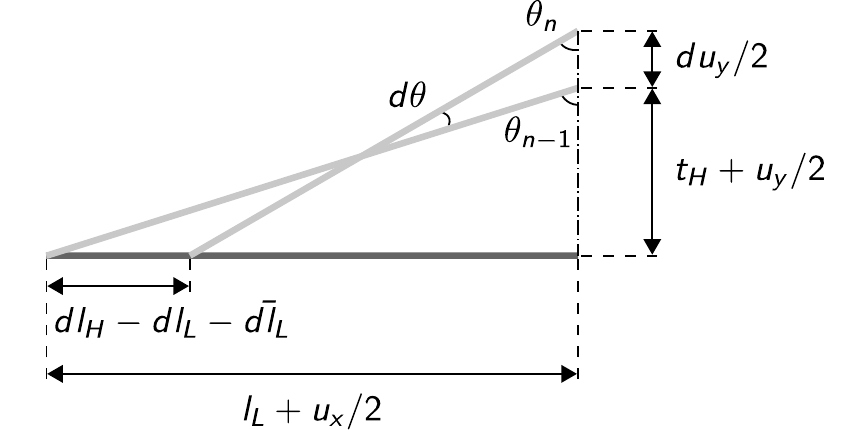}
\caption{Schematic illustrating a quarter of our structure and the deformed angle $d\theta$.}
\label{f:AppAngle}
\end{figure}

We first assume that  $d\theta$ is also equal to the increment in slope of the line connecting the end points of beam 3, as illustrated in Fig.~\ref{f:AppAngle}.
Using trigonometry, we can write:
\begin{multline}
d\theta=\theta_{n-1}-\theta_n=\\
\frac{l_L+u_x/2}{t_H+u_y/2}-\frac{l_L+u_x/2-(dl_H-dl_L-\bar{dl_L})}{t_H+u_y/2+du_y/2}\,\,,
\end{multline}
where $u_x$ and $u_y$ are the total horizontal and vertical displacements of the structure at the previous temperature increment, $l_L=l_1/2$ is the low CTE bar half-length and $t_H=t_3-t_2/2-t_4/2$. Assuming $du_y$ small, an assumption that holds if we consider structures with large aspect ratios $l_L/t_H$ and if we consider small temperature increments $\Delta T$, we can write $t_H+u_y/2\approx t_H+u_y/2+du_y/2$. Thus, our angle becomes
\begin{equation}
d\theta=\frac{dl_H-dl_L-\bar{dl_L}}{t_H+u_y/2}\,\,.
\end{equation}

\section{Temperature-dependent CTE}
\label{a:cte}

\begin{figure}[!htb]
\centering
\includegraphics[scale=1]{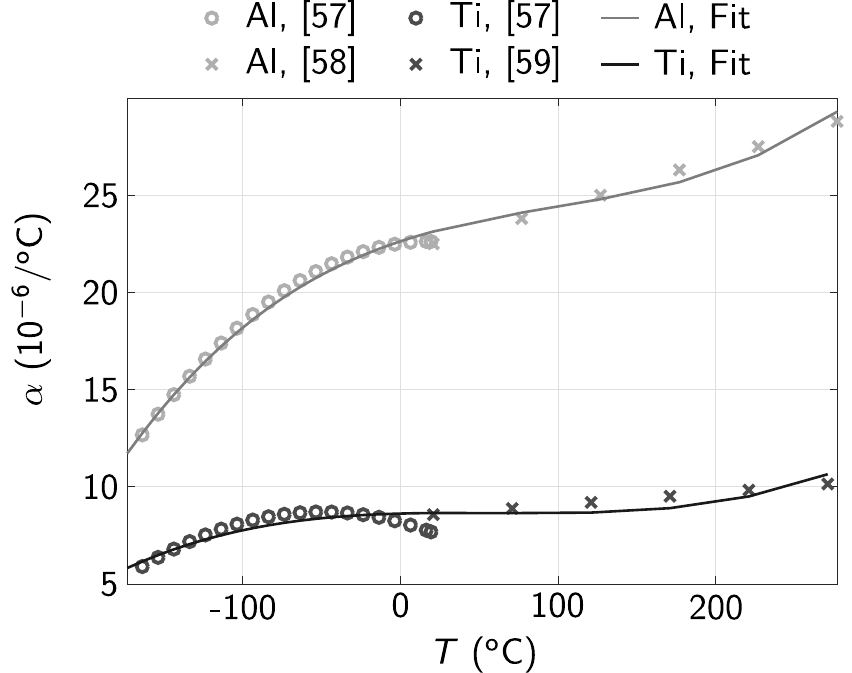}
\caption{Temperature dependence of the coefficients of thermal expansion for Ti and Al, taken from the literature. The continuous lines are obtained by fitting a third order polynomial through the different sets of data for each material.}
\label{f:CteFull}
\end{figure}

Here, we report the temperature-dependent coefficients of thermal expansion for Ti and Al, used in Section~\ref{s:lun} to determine the expansion of one of our structures over the lunar temperature range. These temperature-dependent CTE values for Ti and Al are shown in Fig.~\ref{f:CteFull}. 
For Al, the $\alpha$ at $T<20\,\mathrm{\degree C}$ is obtained from Ref.~\cite{Marquardt2002}, while the values at $T>20\,\mathrm{\degree C}$ are taken from Ref.~\cite{touloukian1979}. For Ti, the $\alpha$ at $T<20\,\mathrm{\degree C}$ is also obtained from Ref.~\cite{Marquardt2002}, while the values at $T>20\,\mathrm{\degree C}$ are derived from Ref.~\cite{Nibennaoune2010}. In our numerical simulations, we use the CTE data obtained by fitting a third order polynomial through the various sets of data we found in the literature for each material. The fitted curves for Ti and Al are shown as continuous lines in Fig.~\ref{f:CteFull}.


\end{document}